\documentclass[amsmath,amssymb,superscriptaddress,twocolumn,aps,prx,longbibliography]{revtex4-2}

\usepackage{empheq}
\usepackage{relsize}
\usepackage{physics}
\usepackage{graphicx}
\usepackage{xcolor}
\usepackage{bm}

\def \a{a}
\def \ad{a^{\dagger}}
\def \Hami{\mathcal{H}}

\renewcommand{\bra}[1]{\langle #1 |}
\renewcommand{\ket}[1]{| #1 \rangle}
\renewcommand{\braket}[2]{\langle #1 | #2 \rangle}

\newcommand{\blue}[1]{\textcolor{blue}{#1}}

\begin{document}

\title{Long-Lived Photon Blockade with Weak Optical Nonlinearity}

\author{You~Wang}
\thanks{These authors contributed equally to this work.}
\affiliation{Division of Physics and Applied Physics, School of Physical and Mathematical Sciences, Nanyang Technological University, Singapore 637371, Singapore}

\author{Xu~Zheng}
\thanks{These authors contributed equally to this work.}
\affiliation{Division of Physics and Applied Physics, School of Physical and Mathematical Sciences, Nanyang Technological University, Singapore 637371, Singapore}

\author{Timothy C.~H.~Liew}
\email{timothyliew@ntu.edu.sg}
\affiliation{Division of Physics and Applied Physics, School of Physical and Mathematical Sciences, Nanyang Technological University, Singapore 637371, Singapore}
\affiliation{MajuLab, International Joint Research Unit UMI 3654, CNRS, Universit\'e C\^ote d'Azur, Sorbonne Universit\'e, National University of Singapore, Nanyang Technological University, Singapore}

\author{Y.~D.~Chong}
\email{yidong@ntu.edu.sg}
\affiliation{Division of Physics and Applied Physics, School of Physical and Mathematical Sciences, Nanyang Technological University, Singapore 637371, Singapore}
\affiliation{Centre for Disruptive Photonic Technologies, School of Physical and Mathematical Sciences, Nanyang Technological University, Singapore 637371, Singapore}

\begin{abstract}
  In conventional photon blockade, the occupation of a cavity mode by more than one photon is suppressed via strong optical nonlinearity.  An alternative, called unconventional photon blockade, can occur under weak nonlinearity by relying on quantum interference between fine-tuned cavities.  A serious limitation is the very short antibunching time window, orders of magnitude less than the cavity lifetime.  We present a method to achieve photon blockade over a large time window of several cavity lifetimes, even exceeding that of conventional photon blockade, while still requiring only weak nonlinearity.  This ``long-lived photon blockade'' (LLPB) occurs when the single-particle Green's function exhibits a zero at a large cavity loss rate, which is satisfied by an exemplary configuration of four coupled cavities under weak driving.  Our analytical results agree well with wavefunction Monte Carlo simulations.  The LLPB phenomenon may aid the development of single-photon sources utilizing materials with weak optical nonlinearities.
\end{abstract}

\maketitle

\textit{Introduction}---Photon blockade is a quantum effect whereby one photon in a nonlinear resonator blocks the entry of other photons, giving rise to antibunched photon statistics~\cite{Tian92,Leo94,Imamo97,Lodahl15}.  Conventionally, photon blockade requires the optical nonlinearity in a cavity to exceed its loss: i.e., $\alpha \gg \gamma$, where $\alpha$ characterizes the strength of the nonlinearity and $\gamma$ is the cavity decay rate.  This can be interpreted as a two-photon state being driven off-resonance by the nonlinearity, shifting its energy by more than the cavity linewidth \cite{Birnbaum05, Dayan08, Hamsen17, Michler00, Faraon08, Claudon10, Ferretti12, He13, Madsen14, Gschrey15, Somaschi16, Dory17, Jia18, Lang11, Wang16, Rabl11, Xu16, Lemonde16, Majumdar12, Majumdar13}.  Achieving this in the optical regime (including telecom wavelengths) is challenging as optical nonlinearities tend to be weak relative to cavity linewidths.  Systems with exceptionally strong nonlinearities are required, such as exciton-polaritons~\cite{Delteil19, Munoz19}, trions~\cite{Kyriienko20}, Rydberg excitons~\cite{Orfanakis22}, and Moire excitons~\cite{Zhang21}.

The strong-nonlinearity condition may be bypassed via the unconventional photon blockade (UPB) phenomenon, which uses destructive interference
between two or more nonlinear cavities (or cavity modes) to cancel the two-photon amplitudes~\cite{Liew10, Ferretti10, Bamba11, Bamba11_2, Flayac13, Xu14, Lemonde14, Gerace14, Flayac15,Shen14,Shen15, Flayac17, Wang17, Sarma17, Ghosh18, Shen18, Sarma18, Ghosh19, Carmichael91, Radulaski17, Kamide17, Zubizarreta20, Wang21}.  It is commonly believed that UPB requires the inter-cavity coupling rate $J$ to satisfy
\begin{equation}
  J \gg \gamma.
  \label{strong-coupling}
\end{equation}
The intuition is that for interference to play a role, the photons should be able to hop many times before leaking away \cite{Liew10}.  In this regime, it turns out that even a weak nonlinearity, $\alpha \ll \gamma$, can cause photon antibunching, i.e., the vanishing of the second-order photon correlation $g^{(2)}(\tau)$ at $\tau = 0$.  However, the suppression only holds over a time window
\begin{equation}
  \delta \tau \sim 1/J \ll 1/\gamma,
\end{equation}
due to the Rabi-like state amplitude oscillations induced by $J$.  This is much smaller than the antibunching time window of $\sim 1/\gamma$ for conventional photon blockade, and poses a serious practical obstacle to the realization of UPB.  For example, a proposal for UPB at telecom wavelengths~\cite{Flayac15} requires $J\approx 20\gamma$ and yields a time window of $\delta \tau\sim100\,\text{ps}$, which is near the resolution limit of current photodetectors.  In two-cavity systems, the UPB condition can be reduced to $\gamma^3\propto\alpha J^2$, and the required cavity quality factors are $Q^3 \propto \omega_0^3 \delta\tau^2 / \alpha$, where $\omega_0$ is the cavity resonance frequency~\cite{Bamba11}.  With a lower bound on $\delta\tau$ imposed by detector resolution limits, this sets a minimum requirement on $Q$ that is unfavorable in the optical regime, where $\omega_0$ is large and optical nonlinearities are typically weak.  Flayac and Savona have suggested that one-way dissipative inter-cavity couplings can suppress the rapid oscillations in $g^{(2)}(\tau)$ \cite{Flayac16}, but such couplings are very challenging to implement experimentally.

In this paper, we show that interference-assisted photon blockade does \textit{not} inherently require the strong-coupling condition \eqref{strong-coupling}, even if we use ordinary couplings.  \blue{In systems of weakly-nonlinear and weakly-driven cavities \cite{Bamba11}, photon blockade emerges perturbatively from linear limit solutions of the system that we call ``single-particle dark state'' (SPDS), defined by the vanishing of the single-particle Green's function in a designated signal cavity.  In the two-cavity systems where UPB was first discovered \cite{Liew10, Ferretti10, Bamba11}, as well as other cavity configurations \cite{Wang21}, it so happens that the SPDS only exists under zero loss ($\gamma = 0$), so the antibunching solutions generated perturbatively from them satisfy Eq.~\eqref{strong-coupling}.  However, the intuition behind Eq.~\eqref{strong-coupling} is not reliable, as there is no general principle that forbids destructive interference from occurring in lossy systems.  As an example, we design a system of four lossy cavities ($\gamma \gg J$) hosting an SPDS, via the exact cancellation of multiple strongly-attenuated interference pathways.}  In the weakly-nonlinear quantum regime ($\alpha \ll \gamma$), this gives rise to a qualitatively different form of antibunching we call long-lived photon blockade (LLPB).  The two-photon correlation function $g^{(2)}(\tau)$ is suppressed over a time window $\delta\tau \approx 8 / \gamma$, two orders of magnitude wider than a comparable two-cavity system exhibiting UPB, and even larger than for conventional photon blockade under strong nonlinearity.  Notably, our setup uses only couplings of the standard (reciprocal and Hermitian) type.  Using these ideas, it may be possible to achieve photon blockade using photonic devices with weak optical nonlinearities.

\textit{Model}---Consider a general system of coupled nonlinear cavities, indexed by $i = 1,\dots$, under a weak coherent drive.  Assuming for simplicity that all the cavities have identical parameters, the cavity Hamiltonian in the frame co-rotating with the driving field, excluding losses, is
\begin{equation}
  \Hami_{0}=\sum_{i\neq j} J_{ij} \ad_i a_j+\sum_i \Delta\ad_i a_i+\alpha\sum_{i}\ad_i\ad_i a_i a_i.
  \label{H0}
\end{equation}
Here, $\hbar = 1$, $\Delta$ is the cavity detuning relative to the driving frequency, $\alpha$ is the nonlinearity strength (Kerr coefficient), $J_{ij}$ is the coupling between cavities $i$ and $j$, and $a_i$ is the photon annihilation operator for cavity $i$. There is no magneto-optic activity, so $J_{ij} = J_{ji} \in \mathbb{R}$ in a suitable gauge.  

We assume a single-input, single-output setup; a single cavity $d$ is driven, with driving Hamiltonian $\Hami_d=F_d\a_d^\dagger+F^*_d a_d$ where $F_d$ is the complex driving amplitude.  (The following derivation can be straightforwardly generalized to multiple coherently driven cavities.)  The density matrix $\rho$ is governed by the Lindblad equation~\cite{BreuerBookOpen}
\begin{equation}
  i\frac{d\rho}{dt}=[\Hami_{0}+\Hami_d,\rho]+\frac{i\gamma}{2}\sum_{i}(2a_i\rho\ad_i-\ad_i a_i\rho-\rho\ad_ia_i),\label{eq:lind}
\end{equation}
where the terms proportional to the cavity loss rate $\gamma$ describe environmental noise.  In the weak drive limit~\cite{Bamba11}, stochastic quantum jumps---the first term in parentheses---can be ignored, and the problem reduces to a dissipative Schr\"odinger equation,
\begin{align}
  i\frac{d|\psi\rangle}{dt} &= \big[\Hami + \Hami_d\big] |\psi\rangle,
  \label{dissipative_schrod} \\
  \Hami
  &\equiv \sum_{i\neq j} J_{ij} \ad_i a_j+\sum_i z\ad_i a_i+\alpha\sum_{i}\ad_i\ad_i a_i a_i, \label{eq:Hamz}\\
  z &\equiv \Delta-i\gamma/2, \label{eq:zdef}
\end{align}
where the non-Hermitian effective Hamiltonian $\Hami$ is derived from \eqref{H0} by adding loss to each site.

The second-order correlation function between sites $i$ and $j$, with time delay $\tau\ge0$, is defined as
\begin{equation}
  g^{(2)}_{ij}(\tau)=\lim\limits_{t\rightarrow + \infty} \frac{\langle\ad_j(t)\ad_i(t+\tau)\a_i(t+\tau)\a_j(t)\rangle}{\langle\ad_j(t)\a_j(t)\rangle\langle\ad_i(t+\tau)\a_i(t+\tau)\rangle},
  \label{g2def}
\end{equation} 
where the expectation values are taken over the vacuum state.  For photon blockade, we are only concerned with self-correlation ($i = j$), but we leave $i, j$  arbitrary for now.  To derive \eqref{g2def}, we develop the perturbative steady-state solution for \eqref{dissipative_schrod} to first order in $\alpha$.  The derivation is presented in the Supplemental Materials~\cite{SM}, and uses the eigendecomposition of the coupling matrix $[J_{ij}] \ket{\varphi_n} = \epsilon_n \ket{\varphi_n}$, where each $|\varphi_n\rangle$ is a single-particle eigenmode and $\epsilon_n$ is the corresponding eigenenergy relative to the driving frequency.  We also define the two-photon basis $\ket{\varphi_{mn}} = \ket{\varphi_m} \otimes \ket{\varphi_n}$.  The result is
\begin{align}
  g^{(2)}_{ij}(\tau)&= \left|1 -  e^{-iz\tau} \, \frac{2\alpha}{G_{id}G_{jd}}
  \sum_k G_{kd}^2 G^{(2)}_{ijkk}(\tau) \right|^2, \label{eq:g2ij_GF}\\
  G_{ij}
  &\equiv\sum_n\frac{\braket{i}{\varphi_n}\braket{\varphi_n}{j}}{z+\epsilon_n}, \label{eq:Gij}\\
  G^{(2)}_{ijkk}(\tau)&\equiv \sum_{mn}
  \frac{\langle i,j|\varphi_{mn}\rangle
    \langle\varphi_{mn}| k,k\rangle}{2z+\epsilon_m+\epsilon_n}e^{-i\epsilon_n\tau}.
  \label{eq:G2tau}
\end{align} 
Here, $|i \rangle \equiv a_i^\dagger|\varnothing \rangle$ is the state with one photon in cavity $i$; $|i,j\rangle\equiv |i\rangle\otimes|j\rangle$;
and $G\equiv\sum_n \ket{\varphi_n}\bra{\varphi_n}/(z+\epsilon_n)$ is the single-particle Green's function for the \textit{linear} system.

The time-dependent correlation function \eqref{eq:g2ij_GF} applies to any configuration of coupled cavities with weak nonlinearity and weak driving, and has not been previously reported to our knowledge.  \blue{From it, we see that the $\tau$-dependence of the correlation has two sources.  The first is the $e^{-iz\tau}$ factor, which simply accounts for the intrinsic frequency detuning and decay of the cavities; if $\Delta$ is small (i.e., the cavities are driven close to resonance), this reduces to $\approx e^{-\gamma\tau/2}$.  This $\tau$-dependence is consistent with the conventional photon blockade result $g^{(2)}(\tau)\approx (1-e^{-\gamma\tau/2})^2$~\cite{SM} (note, however, that conventional blockade uses strong nonlinearity, $\alpha\gg \gamma$).}

\blue{The remaining $\tau$-dependence in \eqref{eq:g2ij_GF} is from $G^{(2)}_{ijkk}(\tau)$, representing interferometric effects.  In Eq.~\eqref{eq:G2tau}, this is written as a superposition of Rabi-like oscillations with eigenfrequencies $\epsilon_n$.  The photon antibunching time-window is therefore determined by $\max \{|\epsilon_n|\}$.}

\begin{figure*}
  \centering
  \includegraphics[width=\linewidth]{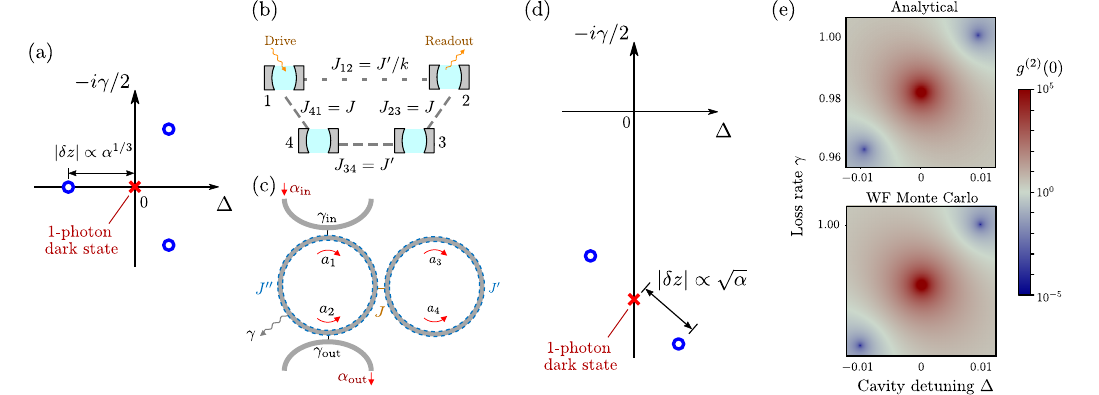}
  \caption{(a) Schematic of zeros (circles) and pole (cross) of $f_{11}(z)$ in the complex $z$ plane, for a two-cavity model.  UPB occurs at the zeros.  (b) Schematic of a four-cavity system exhibiting long-lived photon blockade (LLPB).  (c) Schematic of a photonic structure that can implement the model in (b), using two coupled microring resonators with clockwise and counterclockwise modes of each ring.  (d) Distribution of zeros (circles) and pole (cross) of $f_{22}(z)$ for the four-cavity model of (b).  This illustration is not drawn to scale.  (e) Equal-time two-photon correlation function $g^{(2)}(0)$ for the four-cavity system versus cavity detuning $\Delta$ and loss rate $\gamma$.  Upper plot: analytical results obtained from Eqs.~\eqref{eq:fz_zeros}--\eqref{eq:delta_approx}.  Near the pole (central red spot), there are two zeros of $g^{(2)}(0)$ (blue spots) corresponding to LLPB.  The model parameters are $k = 16$, $J = 0.1227$, $J' = 0.02454$, and $\alpha = 0.001227$, chosen so that one of the LLPB points occurs at $\gamma = 1$.  Lower plot: WFMC simulation results, obtained by directly solving Lindblad master equation with the same parameters and $F_d=10^{-5}$. }
  \label{fig:lattice}
\end{figure*}

\textit{UPB with two cavities}---To illustrate the use of this formalism, consider the two-cavity UPB model with $J_{12} = J_{21} = J \in \mathbb{R}^+$, for which the eigenmodes are $|\varphi_{1,2}\rangle = (|1\rangle \pm |2\rangle)/\sqrt{2}$, with $\epsilon_{1,2} = \pm J$.  Plugging these into Eqs.~\eqref{eq:g2ij_GF}--\eqref{eq:G2tau}, we find that the antibunching condition $g_{11}^{(2)}(0) = 0$ is satisfied if and only if $z^3 \approx -\alpha J^2/2$ \cite{SM}.  In Fig.~\ref{fig:lattice}(a), the three roots for $z$ are plotted as blue circles.  They are symmetrically distributed around the origin (marked by a red cross), the significance of which will be explained shortly.  We focus on the root with positive cavity loss, for which
\begin{equation}
  \gamma=\sqrt{3}\Delta\propto J\sqrt{\alpha/\gamma}.
  \label{eq:upb2_regime}
\end{equation}
Unlike conventional photon blockade, this is achievable for $\alpha \ll \gamma$, but only in the strong-coupling regime \eqref{strong-coupling} \cite{Liew10, Bamba11}.  Moreover, from Eqs.~\eqref{eq:g2ij_GF}--\eqref{eq:G2tau} we can derive
\begin{align}  g_{11}^{(2)}(\tau)
  &\approx \left(1 - e^{-\frac{\gamma}{2}\tau} \cos(\Delta\tau)
  \cos(J\tau)\right)^2,
  \label{eq:g211tau_original}
\end{align}
yielding a short time window of $\delta \tau \sim 1/J$.

\textit{Single-particle dark state}---The condition for photon antibunching in a designated signal cavity $s$ is $g^{(2)}_{ss}(0)=0$. Using Eq.~\eqref{eq:g2ij_GF}, this can be re-expressed as
\begin{equation}
  f_{ss}(z) \equiv \frac{G_{sd}^2-2\alpha\sum_k G_{kd}^2G^{(2)}_{sskk}(0)}{G_{sd}^2}
  = 0.
  \label{eq:fz}
\end{equation}
The Green's functions, and hence $f_{ss}(z)$, are meromorphic functions in the complex $z$ plane.  \blue{We claim that for small $\alpha$, the zeros of $f_{ss}(z)$ lie close to the zeros of $G_{sd}(z)$.  To see why, observe that $f_{ss}(z) = 1$ over the whole $z$-plane when $\alpha = 0$.  For $\alpha > 0$, $f_{ss}$ develops poles at discrete points $\{z_n\}$ which are the zeros of $G_{sd}(z)$.  By Cauchy's argument principle, the net number of zeros and poles (counted with their multiplicities and orders) near each $z_n$ is conserved, so one or more zeros of $f_{ss}(z)$ must emerge continuously from $z_n$.  This completes the argument.}

The zeros of $G_{sd}(z)$, which we call SPDSs, are instances where the \textit{linear} version of the system experiences complete destructive interference in the signal cavity.  Motivated by the above discussion, we seek a system with an SPDS in the  $\gamma \gg \max\{|\epsilon_n|\}$ regime.  Since $\max\{|\epsilon_n|\}$ is the spectral radius of $[J_{ij}]$, which for an $N\times N$ matrix satisfies $\max\{J_{ij}\}\leq\max\{\abs{\epsilon_n}\}\leq N\max\{J_{ij}\}$, \blue{we can reformulate the criterion as
\begin{equation}
  G_{sd}(z) = 0\;\; \mathrm{for}  \;
  -\Im[z] \gg \max\{J_{ij}\}.
  \label{llpb_condition}
\end{equation}
}

\textit{Long-Lived Photon Blockade}---Fig.~\ref{fig:lattice}(b) shows an exemplary four-cavity setup that can satisfy \eqref{llpb_condition}.  The cavities are arranged in a ring with couplings $J_{12}, J_{23}, J_{34}, J_{41} \in \mathbb{R}^+$. The driven cavity is $d = 1$, and the signal cavity is $s=2$.

\blue{To understand intuitively how the SPDS arises, we use a Dyson series in the variable $1/z$, which is the Green's function for an isolated cavity.  Each term in the series can be interpreted as a photon trajectory starting from $d$ and ending at $s$~\cite{Clerk22}; since we intend $\gamma$ to be large relative to the coupling rates, long trajectories that hop multiple times are negligible.  The Dyson series for $G_{21}$ to order $1/z^4$ contains the following five trajectories:
\begin{align}
  &\textbf{Trajectory} &\textbf{Dyson series term} \nonumber\\ 
  &1\mapsto 2 &\frac{1}{z}\cdot J_{12} \cdot \frac{1}{z}\nonumber\\
  &1\mapsto 2\mapsto 1\mapsto 2 &\frac{1}{z}\cdot J_{12} \cdot \frac{1}{z} \cdot J_{12} \cdot \frac{1}{z} \cdot J_{12} \cdot \frac{1}{z}\nonumber\\
  &1\mapsto 2\mapsto 3\mapsto 2 &\frac{1}{z}\cdot J_{12} \cdot \frac{1}{z} \cdot J_{23} \cdot \frac{1}{z} \cdot J_{23} \cdot \frac{1}{z}\nonumber\\
  &1\mapsto 4\mapsto 1\mapsto 2 &\frac{1}{z}\cdot J_{41} \cdot \frac{1}{z} \cdot J_{41} \cdot \frac{1}{z} \cdot J_{12} \cdot \frac{1}{z}\nonumber\\
  &1\mapsto 4\mapsto 3\mapsto 2 &\frac{1}{z}\cdot J_{41} \cdot \frac{1}{z} \cdot J_{34} \cdot \frac{1}{z} \cdot J_{23} \cdot \frac{1}{z}\nonumber
\end{align}}
Equating the sum of these terms to zero, we obtain \cite{SM}
\begin{equation}
  z=\pm i\sqrt{J_{41}^2+J_{12}^2+J_{23}^2+\frac{J_{41}J_{23}J_{34}}{J_{12}}}.
\end{equation}
By choosing $J_{12} \ll J_{14}, J_{23}, J_{34}$, we can get a root
with $\gamma = -2\mathrm{Im}(z) \approx 2(J_{41} J_{23} J_{34} / J_{12})^{1/2}$, satisfying Eq.~\eqref{llpb_condition}.  

\blue{This four-cavity setup appears to be a minimal configuration for achieving LLPB.  Applying the above Dyson series analysis (up to second-lowest order in $1/z$) to two- and three-cavity systems, we find no configurations supporting an SPDS at large $\gamma$. For two cavities, the Dyson series yields $z=\pm iJ_{12}$, whereas for three cavities it gives $z=-J_{12}-J_{13}$ or $z=-J_{13}J_{23}/J_{12}$; none of these solutions satisfy the criterion \eqref{llpb_condition} \cite{SM}.}

We now adopt a mirror-symmetric configuration $J_{12} = J'/k$, $J_{14} = J_{23} = J$, and $J_{34} = J'$.  The above condition for the SPDS can then be met by choosing $J \gtrsim J'$ and $k \gg 1$.  One potential advantage is that this can be realized with the structure shown in Fig.~\ref{fig:lattice}(c), consisting of two ring resonators each hosting a pair of weakly-coupled clockwise and counterclockwise modes~\cite{Little97,Li20,Li12,Morichetti10}.  In the Supplemental Materials, we present further details about this coupled-ring setup, including full-wave simulations showing how to achieve the desired model parameters \cite{SM}.  If we further take $J \gg J'$, we can obtain the closed-form approximation
\begin{align}
  f_{22}(z)&\approx \frac{(z-z_0-\delta z)(z-z_0+\delta z)}{(z-z_0)^2},
  \label{eq:fz_zeros} \\
  z_0 \approx& -i\sqrt{k}J,\hspace{10pt}
  \delta z %
  \approx \pm{\frac{\sqrt{38}}{16}}\sqrt{\alpha \gamma}e^{-i\pi/4}. \label{eq:delta_approx}
\end{align}
The SPDS occurs at $z=z_0$, with $|\mathrm{Im}(z_0)| = \sqrt{k}J \gg J$, thus satisfying Eq.~\eqref{llpb_condition}.  This point, along with the nearby zeros of $f_{22}(z)$---where LLPB is predicted to occur---are depicted in Fig.~\ref{fig:lattice}(d).

In the upper panel of Fig.~\ref{fig:lattice}(e), we plot $g_{22}^{(2)}(0)$ against $\Delta$ and $\gamma$, calculated from the crude approximation Eqs.~\eqref{eq:fz_zeros}--\eqref{eq:delta_approx} using the system parameters stated in the figure caption.  Note that the vertical axis is flipped relative to Fig.~\ref{fig:lattice}(d).  We observe strong local minima, consistent with the above predictions.  For comparison, the lower panel of Fig.~\ref{fig:lattice}(e) shows results from wavefunction Monte Carlo (WFMC)~\cite{Dum92,Molmer93,carmichael_1993,Barchielli_1991} simulations that directly solve the Lindblad equation \eqref{eq:lind} for the same cavity parameters, using $F_d=10^{-5}$ and Fock cutoffs chosen per site to balance accuracy and efficiency. Details about the simulations may be found in the Supplemental Materials~\cite{SM}. One notable difference is that the WFMC simulations have $g_{22}^{(2)}(0) \ll 1$ but not exactly zero; this residual value increases with $F_d$, alongside an increase in the mean photon occupation number in the signal cavity \cite{SM}.

\begin{figure}
  \centering
  \includegraphics[width=0.85\linewidth]{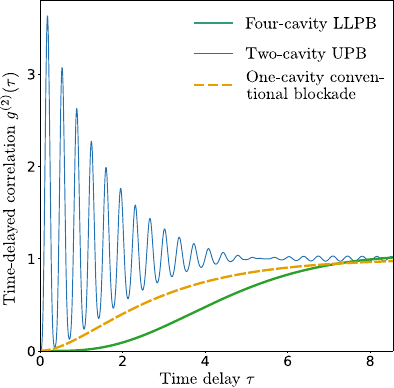}
  \caption{Time-delayed second-order correlation function $g^{(2)}(\tau)$ for a four-cavity system with LLPB (green curve), a two-cavity system with UPB (blue curve), and a one-cavity system with conventional photon blockade (orange dashes).  All results are obtained using WFMC simulations.  The model parameters for the four-cavity LLPB setup are the same as in Fig.~\ref{fig:lattice}, with $\Delta = 0.009571$ and $\gamma = 1$.  The other two cases are tuned for photon blockade at the same $\gamma$.  For the two-cavity system, we choose the same nonlinearity strength $\alpha = 0.001227$, for which UPB occurs at $J \approx 17.67$ and $\Delta \approx 0.2915$. For the one-cavity system, we set $\alpha = 10$ and $\Delta = 0.02491$.}
  \label{fig:g2tau}
\end{figure}

\textit{Antibunching time window}---In Fig.~\ref{fig:g2tau}, we plot the time-delayed correlation function $g^{(2)}(\tau)$ versus $\tau$, for the above four-cavity system at the $\gamma = 1$ LLPB point (green curve).  
The photon antibunching time window, which we characterize by $g^{(2)}(\tau)<0.5$, is $\delta \tau \approx 8/\gamma$.  This plot is obtained from WFMC simulations, but the theory based on Eqs.~\eqref{eq:g2ij_GF}--\eqref{eq:G2tau} gives almost identical results \cite{SM}.
For comparison, the blue curve in Fig.~\ref{fig:g2tau} shows $g^{(2)}(\tau)$ for an analogous two-cavity UPB setup \cite{Liew10}.  With the same nonlinearity strength $\alpha$ and decay rate $\gamma$ used in the four-cavity LLPB result, the two-cavity system achieves UPB for $J \approx 17.67$.  This results in rapid oscillations of $g^{(2)}(\tau)$, with an antibunching window (estimated in the same way) around 90 times smaller than under LLPB.

As a further comparison, the orange dashes in Fig.~\ref{fig:g2tau} show $g^{(2)}(\tau)$ for a one-cavity setup with $\gamma = 1$ and $\alpha = 10$.  (Note that $g^{(2)}(0)$ does not reach precisely zero under conventional photon blockade; we simply choose $\alpha$ so that it is visibly suppressed.)

Returning to the LLPB result, we also note that $g^{(2)}_{22}(\tau)$ scales very slowly for small $\tau$.  To quantify this, we expand it as a polynomial in $\gamma \tau$ for $\gamma \tau \ll 1$.  We obtain
\begin{align}
  g_{22}^{(2)}(\tau)
  &\approx
  (\gamma \tau)^4/64, \label{eq:g2tau_WCUPB_analytical}
\end{align}
which is indeed much slower than the UPB and conventional photon blockade results, both of which scale as $\tau^2$~\cite{SM}.  This ``flat bottom" feature of LLPB may be particularly advantageous for pulsed device operations~\cite{Flayac15,Flayac17}.

\textit{Conclusion}---We have re-analyzed the origins of unconventional photon blockade (UPB), in which photon antibunching is achieved by quantum interference in weakly nonlinear coupled cavities \cite{Liew10}.  A major problem with UPB has hitherto been the rapid oscillation in $g^{(2)}(\tau)$ due to strong inter-cavity coupling \cite{Flayac15, Flayac17}.  We show that this can be overcome by designing a coupled-cavity system so that the linear single-particle Green's function exhibits a zero (an SPDS) at a large cavity loss rate.  This leads to a ``long-lived photon blockade'' (LLPB) scenario whereby $g^{(2)}(\tau) \propto \tau^4$ and the time window exceeds that of conventional photon blockade under strong nonlinearity. It is remarkable that \textit{shortening} the cavity lifetime can effectively \textit{lengthen} the antibunching time window, which can be regarded as a fresh example of non-Hermiticity (in the form of dissipation) producing unexpected physical effects \cite{bender2007making, ashida2020non}.  Our proposal does not need one-way inter-cavity couplings or other exotic requirements~\cite{Flayac16, Flayac17}, and should be compatible with quantum optics experiments using setups like the one shown in Fig.~\ref{fig:lattice}(c) \cite{SM}.

This theoretical framework offers ample scope for further generalization and optimization.  For instance, in the four-cavity configuration of Fig.~\ref{fig:lattice}(b), the couplings are described by just three parameters for simplicity, but they can instead be individually optimized to achieve an even larger $g^{(2)}(\tau)$ time window.  Coherently driving multiple cavities also offers additional possibilities for achieving LLPB with other cavity configurations.  Possible implementation challenges may include the need for fine-tuning and ensuring the stability of the inter-cavity couplings and other system parameters, which is a problem shared with the original UPB setup and all other related schemes to date.

\bibliography{ref}

\clearpage

\begin{widetext}

\makeatletter 
\renewcommand{\theequation}{S\arabic{equation}}
\makeatother
\setcounter{equation}{0}

\makeatletter 
\renewcommand{\thefigure}{S\@arabic\c@figure}
\makeatother
\setcounter{figure}{0}

\makeatletter 
\renewcommand{\thesection}{S\arabic{section}}
\makeatother
\setcounter{section}{0}

\setcounter{page}{1}

\begin{center}
  {\Large Supplemental Materials}
\end{center}
  
\section{Derivation of delayed second-order correlations under weak driving}\label{app:delayed_g2tau}

As mentioned in the main text, the steady state for a system of driven nonlinear cavities, in the limit where the driving is weak, can be obtained from a dissipative effective Schr\"odinger equation.  Following Ref.~\cite{Bamba11}, under the weak-driving assumption the Hilbert space is divided into subspaces of different photon number, and the wavefunction is written as a superposition of contributions from each subspace:
\begin{equation}
  \ket{\psi}=\ket{\psi^{(0)}}+\ket{\psi^{(1)}}+\ket{\psi^{(2)}}+\hdots
\end{equation}
Under weak driving, the amplitudes in higher photon number subspaces should be negligible.  Hence, we solve the Schr\"{o}dinger equation for each subspace recursively~\cite{Bamba11}, truncating at a given photon number bound:
\begin{align}
  i\frac{d}{dt}\ket{\psi^{(k)}}&=\Hami\ket{\psi^{(k)}}+F_d\ad_d\ket{\psi^{(k-1)}}+F_d^*a_d\ket{\psi^{(k+1)}} \nonumber \\ 
  &\approx\Hami\ket{\psi^{(k)}}+F_d\ad_d\ket{\psi^{(k-1)}}. \label{eq:psik_schordinger}
\end{align}
The steady-state wavefunction $\ket{\bar{\psi}}$ can be found by setting the time derivative in each subspace to zero.  We obtain
\begin{equation}
  \ket{\bar{\psi}^{(k)}}=-F_d\Hami^{-1}\ad_d\ket{\bar{\psi}^{(k-1)}}, \label{eq:psikbar}
\end{equation}
with $\ket{\bar{\psi}^{(0)}}=\ket{\psi^{(0)}}$ being the vacuum state.

For the single-photon subspace, the effective Hamiltonian is $\Hami = [J_{ij}]+z\mathcal{I}$. This subspace is spanned by the eigenvectors of the coupling matrix, i.e.,
\begin{equation}
  [J_{ij}] \ket{\varphi_n} = \epsilon_n \ket{\varphi_n}.
\end{equation}
Supposing that one cavity $d$ is driven, \eqref{eq:psikbar} reduces to
\begin{equation}
  \ket{\bar{\psi}^{(1)}}=-F_dG\ket{d},\quad \textrm{ where } G\equiv\Hami^{-1}=\sum_n\frac{\ket{\varphi_n}\bra{\varphi_n}} {z+\epsilon_n}.
\end{equation}

In this work, we truncate the calculation at two photons.  For the 2-photon subspace, we use the basis formed by tensor products $\ket{\varphi_{mn}} = \ket{\varphi_m} \otimes \ket{\varphi_n}$.  We then calculate the steady-state amplitude in this subspace using a perturbation expansion to first order in the nonlinearity strength $\alpha$~\cite{Wang21}:
\begin{align}
  \ket{\bar{\psi}^{(2)}}&\approx\ket{\bar{\psi}_0^{(2)}}+\alpha\ket{\bar{\psi}_1^{(2)}} \nonumber\\
  &=\frac{1}{\sqrt{2}}\ket{\bar\psi^{(1)}}\otimes \ket{\bar\psi^{(1)}}-\sqrt{2}\alpha F_d^2\sum_k G_{kd}^2G^{(2)}\ket{k,k}, \label{eq:psi2bar}
\end{align}
where $G_{kd}\equiv\bra{k}G\ket{d}$ is the $k,d$ element of the single-particle Green's function and 
\begin{equation}
  G^{(2)}\equiv\sum_{mn}\frac{\ket{\varphi_{nm}}\bra{\varphi_{mn}}}{2z+\epsilon_m+\epsilon_n} \label{eq:G2}
\end{equation}
is the linear two-photon Green's function.

The second-order correlation function between sites $i$ and $j$, with time delay $\tau\ge0$, is defined as
\begin{equation}
  g^{(2)}_{ij}(\tau)=\lim\limits_{t\rightarrow + \infty} \frac{\langle\ad_j(t)\ad_i(t+\tau)\a_i(t+\tau)\a_j(t)\rangle}{\langle\ad_j(t)\a_j(t)\rangle\langle\ad_i(t+\tau)\a_i(t+\tau)\rangle},
\end{equation} 
where the expectation values are taken over the vacuum state.  In the Schr\"{o}dinger representation, this can be transformed to \cite{Shen14}
\begin{align}
  g^{(2)}_{ij}(\tau)&=\frac{\mathrm{tr}[\bar{\rho}\ad_j U^{\dagger}(\tau)\ad_i\a_iU(\tau)\a_j]}{\bar{n}_i\bar{n}_j}\nonumber\\
  &=\frac{\mathrm{tr}[U(\tau)\rho_j(0)U^\dagger(\tau)\ad_i\a_i]}{\bar{n}_i\bar{n}_j}, \label{eq:g2ijtau}
\end{align}
where $U(\tau)=\mathrm{exp}(-i\Hami_{tot}\tau)$ is the time-evolution operator, $\bar{\rho}$ is the steady state density operator, and $\rho_j(0)\equiv\a_j\bar{\rho}\ad_j$. In the denominator, $\bar{n}_i\approx |\braket{i}{\bar{\psi}^{(1)}}|^2=|F_dG_{id}|^2$ under the weak drive approximation.  The numerator can be calculated as $\mathrm{tr}[\rho_j(\tau)\ad_i\a_i]$, where
\begin{equation}
  \begin{cases}
    \rho_j(0)=\a_j\bar{\rho}\ad_j\\
    \\
    \rho_j(\tau)=U(\tau)\rho_j(0)U^\dagger(\tau). \label{eq:rhojtau}
  \end{cases}
\end{equation}
This is equivalent to calculating the $i$-th site's photon population at time $\tau$, using the modified initial state $\rho_j(0)$.

Now we convert the density operator back into the form $\rho_j(\tau)\equiv\ket{\psi_j(\tau)}\bra{\psi_j(\tau)}$, and plug it into the numerator in Eq.~\eqref{eq:g2ijtau}:
\begin{equation}
  \mathrm{tr}[U(\tau)\rho_j(0)U^\dagger(\tau)\ad_i\a_i]=\mathrm{tr}[\rho_j(\tau)\ad_i\a_i]\approx |\braket{i}{\psi_j^{(1)}(\tau)}|^2.
\end{equation}
The Schr\"{o}dinger equation corresponding to Eq.~\eqref{eq:rhojtau} is
\begin{equation}
  \begin{cases}
    \ket{\psi_j(0)}=a_j\ket{\bar{\psi}}\\
    \\
    \displaystyle i\frac{d}{d\tau}\ket{\psi_j(\tau)}=\Hami_{tot}\ket{\psi_j(\tau)}. 
  \end{cases} \label{eq:psij_evo}
\end{equation}
We then recursively solve Eq.~\eqref{eq:psij_evo} via Eq.~\eqref{eq:psik_schordinger}:
\begin{align}
  \ket{\psi_j^{(k)}(0)}&=\a_j\ket{\bar{\psi}^{(k+1)}},\nonumber\\
  i\frac{d}{d\tau}\ket{\psi_j^{(k)}(\tau)}&=\Hami\ket{\psi_j^{(k)}(\tau)}+\Hami_+\ket{\psi_j^{(k-1)}(\tau)}.
\end{align}

For the single-photon ($k=1$) subspace,
\begin{align}
  \ket{\psi_j^{(1)}(0)}&=\a_j\ket{\bar{\psi}^{(2)}}, \nonumber\\
  i\frac{d}{d\tau}\ket{\psi_j^{(1)}(\tau)}&=\Hami\ket{\psi_j^{(1)}(\tau)}+F_d\ad_d\a_j\ket{\bar{\psi}^{(1)}} \nonumber\\
  &=\Hami\ket{\psi_j^{(1)}(\tau)}+F_d\braket{j}{\bar{\psi}^{(1)}}\ket{d}, \label{eq:psij1tau_schordinger}
\end{align}
where we make use of the fact that $\ket{\psi_j^{(0)}(\tau)}=\ket{\psi_j^{(0)}(0)}=\a_j\ket{\bar{\psi}^{(1)}}$ is the vacuum state.  The solution to Eq.~\eqref{eq:psij1tau_schordinger} is
\begin{equation}
  \ket{\psi_j^{(1)}(\tau)}=\ket{\psi_j^{(1)}(+\infty)}+e^{-i\Hami\tau}[\ket{\psi_j^{(1)}(0)}-\ket{\psi_j^{(1)}(+\infty)}]. \label{eq:psij1tau}
\end{equation}
The steady state $\ket{\psi_j^{(1)}(+\infty)}$ can be obtained by setting the time derivative to zero in Eq.~\eqref{eq:psij1tau_schordinger}:
\begin{align}
  0 = & \Hami\ket{\psi_j^{(1)}(+\infty)}+F_d\braket{j}{\bar{\psi}^{(1)}}\ket{d}, \nonumber\\
  \ket{\psi_j^{(1)}(+\infty)}&=-\Hami^{-1}F_d\braket{j}{\bar{\psi}^{(1)}}\ket{d}=\braket{j}{\bar{\psi}^{(1)}}\ket{\bar{\psi}^{(1)}}.\label{eq:psij1inf}
\end{align}
The initial state $|\psi_j^{(1)}(0)\rangle = a_j |\bar{\psi}^{(2)}\rangle$ is directly obtained by referring to Eq.~\eqref{eq:psi2bar}:
\begin{align}
  |\psi_j^{(1)}(0)\rangle=a_j|\bar{\psi}^{(2)}\rangle &= \langle j|\bar{\psi}^{(1)}\rangle|\bar{\psi}^{(1)}\rangle - \sqrt{2}\alpha F_d^2 \sum_{mnk} G_{kd}^2 \frac{\langle\varphi_{mn}|k,k\rangle}{2z+\epsilon_m+\epsilon_n}\frac{\langle j|\varphi_m\rangle|\varphi_n\rangle+\langle j|\varphi_n\rangle|\varphi_m\rangle}{\sqrt{2}}, \nonumber\\
  &= \langle j|\bar{\psi}^{(1)}\rangle|\bar{\psi}^{(1)}\rangle - 2\alpha F_d^2 \sum_{mnk} \frac{G_{kd}^2\langle\varphi_{mn}|k,k\rangle\langle j|\varphi_m\rangle}{2z+\epsilon_m+\epsilon_n}|\varphi_n\rangle. \label{eq:psij10}
\end{align}
Finally, plugging Eq.~\eqref{eq:psij10} and Eq.~\eqref{eq:psij1inf} back into Eq.~\eqref{eq:psij1tau} gives
\begin{equation}
    \langle i|\psi_j^{(1)}(\tau)\rangle = \langle j|\bar{\psi}^{(1)}\rangle\langle i|\bar{\psi}^{(1)}\rangle - 2\alpha F_d^2 e^{-iz\tau}\sum_k G_{kd}^2 \langle i,j|G^{(2)}(\tau)|k,k\rangle, \\
\end{equation}
where
\begin{equation}
  G^{(2)}(\tau) \equiv \sum_{mn} \frac{|\varphi_{mn}\rangle\langle\varphi_{mn}|}{2z+\epsilon_m+\epsilon_n} e^{-i\epsilon_n\tau}.
\end{equation}
We finally arrive at a perturbative solution for the delayed second-order correlation function:
\begin{align}
    g_{ij}^{(2)}(\tau) &= \left|\frac{\langle i|\psi_j^{(1)}(\tau)\rangle^2}{\bar{n}_i\bar{n}_j}\right| = \left|\frac{\langle i|\psi_j^{(1)}(\tau)\rangle}{\langle i|\bar{\psi}^{(1)}\rangle\langle j|\bar{\psi}^{(1)}\rangle}\right|^2 \\
  &= \left|1 - 2\alpha e^{-iz\tau}\sum_k \frac{G_{kd}^2}{G_{id}G_{jd}} \langle i,j|G^{(2)}(\tau)|k,k\rangle\right|^2. \label{eq:g2ijtau_GF}
\end{align}

\section{Conventional photon blockade}\label{app:CPB}

To model conventional photon blockade model, we cannot use the perturbative procedure from the previous section, as $\alpha$ is large.  However, Eqs.~\eqref{eq:g2ijtau}--\eqref{eq:psij1tau} remain valid, and can be solved directly for a single-cavity system.  In this case, the non-driven Hamiltonian is
\begin{equation}
  \Hami=z\ad_1 a_1+\alpha\ad_1\ad_1a_1a_1,
\end{equation}
and the 1-photon and 2-photon Green's functions are
\begin{equation}
  G = \frac{|1\rangle\langle1|}{z}, \quad G^{(2)} = \frac{|1,1\rangle\langle1,1|}{2z+2\alpha}.
\end{equation}
According to Eq.~\eqref{eq:psikbar}, the steady state wavefunctions are
\begin{equation}
  |\bar{\psi}^{(1)}\rangle = -\frac{F}{z}|1\rangle, \quad |\bar{\psi}^{(2)}\rangle = \frac{F^2}{\sqrt{2}z(z+\alpha)}|1,1\rangle.
\end{equation}
Plugging into Eqs.~\eqref{eq:psij1tau_schordinger} and Eq.~\eqref{eq:psij1tau}, we obtain
\begin{align}
  |\psi_1^{(1)}(0)\rangle &= a_1|\bar{\psi}^{(2)}\rangle = \frac{F^2}{z(z+\alpha)}|1\rangle \\
  |\psi_1^{(1)}(+\infty)\rangle &= \frac{F^2}{z^2}|1\rangle\\
  \Rightarrow \qquad |\psi_1^{(1)}(\tau)\rangle &= \frac{F^2}{z^2}\left(1-\frac{\alpha}{\alpha+z}e^{-i z\tau}\right)|1\rangle.
\end{align}
Thus, the second-order correlation function is
\begin{equation}
  g_{11}^{(2)}(\tau) = \left|\frac{\langle1|\psi_1^{(1)}(\tau)\rangle}{\langle1|\bar{\psi}^{(1)}\rangle^2}\right|^2 = \left|1-\frac{\alpha}{\alpha+z}e^{-iz\tau}\right|^2.
\end{equation}
Optimal parameters for photon blockade can be found by minimizing the equal-time second-order correlation:
\begin{align}
  g_{11}^{(2)}(0) &= \left|\frac{z}{\alpha+z}\right|^2 = \frac{\Delta^2+\frac{\gamma^2}{4}}{(\Delta+\alpha)^2+\frac{\gamma^2}{4}},\\
  \Delta_{min} &= -\frac{1}{2}(\alpha-\sqrt{\alpha^2+\gamma^2}).
\end{align}
For large nonlinearity $\alpha\gg \gamma$, we can achieve $\Delta_{min}\approx0$, and the corresponding time-delayed correlation function is
\begin{align}  
  g_{11}^{(2)}(\tau) &= 1-\frac{4\alpha^2(2e^{-\frac{\gamma}{2}\tau}-e^{-\gamma\tau})}{4\alpha^2+\gamma^2}\\
  &=\frac{\gamma^2}{4\alpha^2+\gamma^2}+\frac{4\alpha^2}{4\alpha^2+\gamma^2}\left(1-e^{-\frac{\gamma}{2}\tau}\right)^2 \label{g11result}\\
  &\approx \left(1-e^{-\frac{\gamma}{2}\tau}\right)^2.
\end{align}
Fig.~\ref{fig:g2tau_CPB} shows an exemplary plot of $g_{11}^{(2)}(\tau)$ versus $\tau$, comparing Eq.~\eqref{g11result} (solid blue curve) to wavefunction Monte Carlo (WFMC) simulations (black dots).  The theoretical results are clearly very accurate.

\begin{figure}
  \centering
  \includegraphics[width=0.4
  \textwidth]{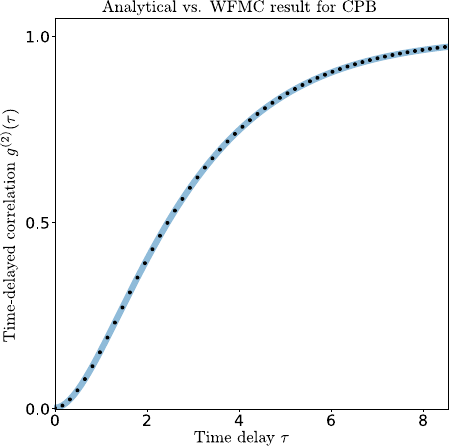}
  \caption{Time-delayed correlation function $g_{11}^{(2)}(\tau)$ for conventional photon blockade in a single cavity, using the analytical result \eqref{g11result} (solid blue line) and WFMC simulations (black dots).  The system parameters are the same as in the conventional blockade system from Fig.~2 of the main text.} 
  \label{fig:g2tau_CPB}
\end{figure}

\section{Unconventional photon blockade in two cavities}
\label{app:originalUPB}

In this section, we re-derive unconventional photon blockade (UPB) in a two-cavity system using the framework of Section~\ref{app:delayed_g2tau}.  Here, the coupling Hamiltonian is
\begin{equation}
  [J_{ij}]=\begin{pmatrix}0&J\\J&0\end{pmatrix},
\end{equation}
and its eigenvectors and eigenvalues are
\begin{align}
  \begin{pmatrix}|\varphi_1\rangle & |\varphi_2\rangle\end{pmatrix} = \frac{1}{\sqrt{2}}\begin{pmatrix} -1 & 1 \\ 1 & 1 \end{pmatrix}, \quad \begin{pmatrix}\epsilon_1 \\ \epsilon_2\end{pmatrix} = \begin{pmatrix} -J \\ J \end{pmatrix}. \label{eq:dimer_eigen}
\end{align}
As in the original UPB model \cite{Liew10, Ferretti10, Bamba11}, we drive and measure at cavity 1.  From Eq.~\eqref{eq:g2ijtau_GF},
\begin{align}
  g_{ij}^{(2)}(\tau) &= \left|1-2\alpha e^{-iz\tau}\sum_n A_{ij}(n)e^{-i\epsilon_n\tau}\right|^2,\label{eq:g2ijtau_rabi1} \\ A_{ij}(n)&=\sum_{km}\frac{G^2_{kd}}{G_{id}G_{jd}}\frac{\langle i,j|\varphi_{mn}\rangle\langle\varphi_{mn}|k,k\rangle}{2z+\epsilon_m+\epsilon_n}.
  \label{eq:g2ijtau_rabi}
\end{align}
By inserting Eq.~\eqref{eq:dimer_eigen} into Eq.~\eqref{eq:g2ijtau_rabi} (note that $G=\sum_n (z+\epsilon_n)^{-1}\ket{\varphi_n}\bra{\varphi_n}$), we have
\begin{align}
  A_{11}(1) &= \frac{2z^3-Jz^2+J^3}{8z^3(z-J)}, \quad A_{11}(2) = -\frac{2z^3+Jz^2-J^3}{8z^3(z+J)}.
\end{align}
Hence the interfered oscillation terms sum up to
\begin{equation}
  \sum_n A_{11}(n)e^{-i\epsilon_n\tau} = \frac{2z^4-J^2z^2+J^4}{4z^3(z^2-J^2)}\cos J \tau + i\frac{Jz(z^2+J^2)}{4z^3(z^2-J^2)}\sin J \tau.
\end{equation}
The optimal antibunching condition is
\begin{equation}
  g_{11}^{(2)}(0) = \left|1-2\alpha\cdot\frac{2z^4-J^2z^2+J^4}{4z^3(z^2-J^2)}\right|^2 = 0.
\end{equation}
This can be satisfied for weak nonlinearity (i.e., small $\alpha$) by taking $J$ to be large, so that
\begin{equation}
  z = \Delta-\frac{i}{2}\gamma \approx \left(\frac{\alpha J^2}{2}\right)^{\frac{1}{3}}e^{-\frac{i\pi}{3}}.
\end{equation}
This is consistent with the results of earlier studies, such as Ref.~\cite{Bamba11}.  The corresponding delayed second-order correlation function is
\begin{align}  
  g_{11}^{(2)}(\tau) &= \left|1-e^{-iz\tau}\cos J \tau+iO(\alpha^{\frac{1}{3}})\right|^2 \nonumber\\
  &\approx \left|1-e^{-\frac{\gamma}{2}\tau}e^{-i\Delta\tau} \cos J \tau\right|^2 \nonumber\\
  &\approx \left(1-e^{-\frac{\gamma}{2}\tau}\cos{\Delta\tau}\cos J \tau\right)^2.
\end{align}
In the last line, we have used the fact that $\Delta\ll J$ since $(\Delta/J)^3\approx (1/16)\alpha/J\ll 1$, and the imaginary part of the slow-varying $e^{-i\Delta\tau}$ can be omitted. In Fig.~\ref{fig:g2tau_UPB}, we compare this analytical formula to WFMC simulation results, using the same UPB settings as in Fig.~2 of the main text.

\begin{figure}
  \centering
  \includegraphics[width=0.4\textwidth]{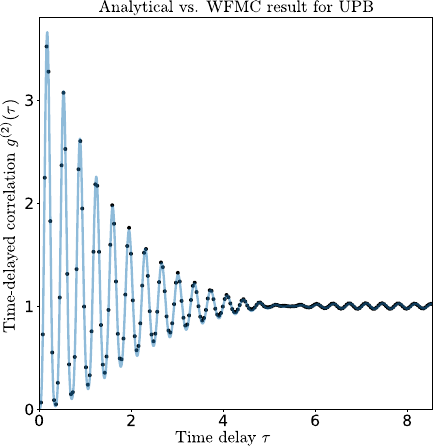}
  \caption{Time-delayed correlation function $g_{11}^{(2)}(\tau)$ for UPB in a two-cavity system, showing analytical results (solid blue curve) and WFMC simulation results (black dots).  The system parameters are the same as in Fig.~2 of the main text.} 
  \label{fig:g2tau_UPB}
\end{figure}

\section{Long-lived photon blockade}
\label{app:GF0entry}
\label{app:weakly coupled UPB}

As discussed in the main text, our desired route to photon blockade is to find a situation where $G_{ij} = 0$ at a large value of the cavity decay rate (relative to the coupling strength).  Let us consider a more general scenario than in the main text, whereby all the couplings $J_{ij}$ and the complex detunings $z_i$ are independently tunable.  For the two-cavity system, we have
\begin{equation}
  \Hami=\begin{pmatrix}
    z_1&J\\J&z_2
  \end{pmatrix},\hspace{10pt}
  G=\frac{1}{z_1z_2-J^2}\begin{pmatrix}
    z_2&-J\\-J&z_1
  \end{pmatrix}.
\end{equation}
Clearly, if $J$ is non-vanishing, a zero of $G$ is only achievable for $z_1=0$ or $z_2=0$, which does not meet our requirements.

For three cavities,
\begin{equation}
  \Hami=\begin{pmatrix}
    z_1&J_1&J_3\\J_1&z_2&J_2\\J_3&J_2&z_3
  \end{pmatrix},\hspace{10pt}
  G=\frac{1}{z_1z_2z_3-z_1J_2^2-z_2J_3^2-z_3J_1^2+2J_1J_2J_3}\begin{pmatrix}
    z_2z_3-J_2^2&J_2J_3-z_3J_1&J_1J_2-z_2J_3\\J_2J_3-z_3J_1&z_3z_1-J_3^2&J_3J_1-z_1J_2\\J_1J_2-z_2J_3&J_3J_1-z_1J_2&z_1z_2-J_1^2
  \end{pmatrix}.
\end{equation}
There are two cases to consider.  (i) To create a zero on the diagonal, two complex detunings must be conjugates of each other (e.g., $z_2=z_3^*$ and $|z_2|=|z_3|=J_2$). This requires either that one cavity has gain precisely balanced with another cavity's loss, or that both cavities are lossless. (ii) To create a zero off-diagonal entry, one complex detuning must be real (e.g., $z_3=J_2J_3/J_1$), implying that the cavity must be completely lossless.  Either requirement is somewhat impractical (though not impossible) compared to the four-cavity setup.

Let us now analyze the four-cavity system, shown schematically in Fig.~1(b) of the main text.

First, we will develop some intuition about how photon antibunching arises in this system.  In the Dyson series expansion for $G_{id}$ as a function of $z$, each term in the series can be interpreted as a photon trajectory starting at the driven cavity $d$ and ending at the output cavity $i$~\cite{Clerk22}.  Along a trajectory, each cavity contributes $1/z$ to the Green's function.  For $G_{21}$, the Dyson series up to order $1/z^4$ has five terms, corresponding to the following trajectories:
\begin{align}
  &\textbf{Trajectory} &\textbf{Dyson series term} \nonumber\\ 
  &1\mapsto 2 &\frac{1}{z}\cdot J_{12} \cdot \frac{1}{z}\nonumber\\
  &1\mapsto 2\mapsto 1\mapsto 2 &\frac{1}{z}\cdot J_{12} \cdot \frac{1}{z} \cdot J_{12} \cdot \frac{1}{z} \cdot J_{12} \cdot \frac{1}{z}\nonumber\\
  &1\mapsto 2\mapsto 3\mapsto 2 &\frac{1}{z}\cdot J_{12} \cdot \frac{1}{z} \cdot J_{23} \cdot \frac{1}{z} \cdot J_{23} \cdot \frac{1}{z}\nonumber\\
  &1\mapsto 4\mapsto 1\mapsto 2 &\frac{1}{z}\cdot J_{41} \cdot \frac{1}{z} \cdot J_{41} \cdot \frac{1}{z} \cdot J_{12} \cdot \frac{1}{z}\nonumber\\
  &1\mapsto 4\mapsto 3\mapsto 2 &\frac{1}{z}\cdot J_{41} \cdot \frac{1}{z} \cdot J_{34} \cdot \frac{1}{z} \cdot J_{23} \cdot \frac{1}{z}\nonumber
\end{align}
If we take the sum of these terms and equating it zero, we arrive at the following solution:
\begin{equation}
  z=\pm i\sqrt{J_{41}^2+J_{12}^2+J_{23}^2+\frac{J_{41}J_{23}J_{34}}{J_{12}}}.
\end{equation}
When $J_{12} \ll J_{41}, J_{23}, J_{34}$, this solution is self-consistent as $\gamma = -2\mathrm{Im}(z) \approx 2\sqrt{J_{41} J_{23} J_{34} / J_{12}}$ is indeed larger than all the coupling terms.

Having established the feasibility of LLPB in this system, let us consider the simplified case $J'/J\ll 1$, which locates the pole at $z_0=-iJ\sqrt{k-J'^2/J^2}\approx -iJ\sqrt{k}$ and $\gamma\approx2\sqrt{k}J$.  Then the eigenvectors and eigenvalues of the coupling Hamiltonian $[J_{ij}]$ can be analytically written as second-order perturbation of $J'$:
\begin{align}
  (\ket{\varphi_1}~\ket{\varphi_2}~\ket{\varphi_3}~\ket{\varphi_4})&=\frac{1}{2}\begin{pmatrix}
    -1&-1&1&1\\
    1&-1&-1&1\\
    -1&1&-1&1\\
    1&1&1&1
  \end{pmatrix}+\frac{J'}{8J}\begin{pmatrix}
    1-\frac{1}{k}+\frac{J^{\prime}}{8J}&-1+\frac{1}{k}+\frac{J^{\prime}}{8J}&1-\frac{1}{k}-\frac{J^{\prime}}{8J}&-1+\frac{1}{k}-\frac{J^{\prime}}{8J}\\
    -1+\frac{1}{k}-\frac{J^{\prime}}{8J}&-1+\frac{1}{k}+\frac{J^{\prime}}{8J}&-1+\frac{1}{k}+\frac{J^{\prime}}{8J}&-1+\frac{1}{k}-\frac{J^{\prime}}{8J}\\
    -1+\frac{1}{k}+\frac{J^{\prime}}{8J}&-1+\frac{1}{k}-\frac{J^{\prime}}{8J}&1-\frac{1}{k}+\frac{J^{\prime}}{8J}&1-\frac{1}{k}-\frac{J^{\prime}}{8J}\\
    1-\frac{1}{k}-\frac{J^{\prime}}{8J}&-1+\frac{1}{k}-\frac{J^{\prime}}{8J}&-1+\frac{1}{k}-\frac{J^{\prime}}{8J}&1-\frac{1}{k}-\frac{J^{\prime}}{8J}
  \end{pmatrix},\\
  \begin{pmatrix}\epsilon_1\\\epsilon_2\\\epsilon_3\\\epsilon_4\end{pmatrix}&=J\begin{pmatrix}
    -1\\-1\\1\\1
  \end{pmatrix}+\frac{J^{\prime}}{2}\begin{pmatrix}
    -1-\frac{1}{k}-\frac{J^{\prime}}{4J}\\1+\frac{1}{k}-\frac{J^{\prime}}{4J}\\-1-\frac{1}{k}+\frac{J^{\prime}}{4J}\\1+\frac{1}{k}+\frac{J^{\prime}}{4J}
  \end{pmatrix}.
\end{align}
The single photon Green's function to the second order of $J^{\prime}/J$ are given by
\begin{align}
  G =\left(
    \begin{array}{cccc}
     \frac{J^2 J^{\prime 2} z}{\left(z^2-J^2\right)^3}+\frac{z}{z^2-J^2} & -\frac{J^{\prime} z^2}{k \left(J^2-z^2\right)^2}-\frac{J^2 J^{\prime}}{\left(J^2-z^2\right)^2} & \frac{J J^{\prime} z}{k \left(J^2-z^2\right)^2}+\frac{J J^{\prime} z}{\left(J^2-z^2\right)^2} & \frac{J J^{\prime 2} z^2}{\left(J^2-z^2\right)^3}+\frac{J}{J^2-z^2} \\
     -\frac{J^{\prime} z^2}{k \left(J^2-z^2\right)^2}-\frac{J^2 J^{\prime}}{\left(J^2-z^2\right)^2} & \frac{J^2 J^{\prime 2} z}{\left(z^2-J^2\right)^3}+\frac{z}{z^2-J^2} & \frac{J J^{\prime 2} z^2}{\left(J^2-z^2\right)^3}+\frac{J}{J^2-z^2} & \frac{J J^{\prime} z}{k \left(J^2-z^2\right)^2}+\frac{J J^{\prime} z}{\left(J^2-z^2\right)^2} \\
     \frac{J J^{\prime} z}{k \left(J^2-z^2\right)^2}+\frac{J J^{\prime} z}{\left(J^2-z^2\right)^2} & \frac{J J^{\prime 2} z^2}{\left(J^2-z^2\right)^3}+\frac{J}{J^2-z^2} & \frac{J^{\prime 2} z^3}{\left(z^2-J^2\right)^3}+\frac{z}{z^2-J^2} & -\frac{J^2 J^{\prime}}{k \left(J^2-z^2\right)^2}-\frac{J^{\prime} z^2}{\left(J^2-z^2\right)^2} \\
     \frac{J J^{\prime 2} z^2}{\left(J^2-z^2\right)^3}+\frac{J}{J^2-z^2} & \frac{J J^{\prime} z}{k \left(J^2-z^2\right)^2}+\frac{J J^{\prime} z}{\left(J^2-z^2\right)^2} & -\frac{J^2 J^{\prime}}{k \left(J^2-z^2\right)^2}-\frac{J^{\prime} z^2}{\left(J^2-z^2\right)^2} & \frac{J^{\prime 2} z^3}{\left(z^2-J^2\right)^3}+\frac{z}{z^2-J^2} \\
    \end{array}.
    \right)
\end{align}
Let $G_{12}=0$, we get the pole $z_0=-i\sqrt{k}J$. Expanding $z$ near the pole as $z=z_0+\delta z$, plugging the eigenvectors, eigenvalues and Green's function into Eq.~\eqref{eq:g2ijtau_rabi}, we get:
\begin{align}
  2\alpha\sum_nA_{22}(n)e^{-i\epsilon_n\tau}\approx&\frac{i\alpha J}{1024 \delta z^2}\left\{\cos (J\tau) \left[16 J\tau \left(-3 k^2-10k\right)-304 \sqrt{k}+3\frac{J^{\prime 2}\tau}{J} k^3\right]\right.\nonumber\\
  &~~~~~~~~~~+\left.\sin (J\tau) \left[48k^2-144k-3 \frac{J^{\prime 2}\tau}{J} k^{5/2} -3\frac{J^{\prime 2}}{J^2} k^3\right]\right\}.
  \label{eq:a22sum}
\end{align}
Requiring $g^{(2)}_{22}(0)=0$ gives
\begin{align}
  -\frac{19i\alpha \sqrt{k}J}{64 \delta z^2}=1\Rightarrow\delta z=\pm\frac{\sqrt{19}}{8}k^{1/4}\sqrt{\alpha J}e^{-\pi i/4}.
\end{align}
Hence, Eq.\eqref{eq:a22sum} becomes
\begin{align}
  2\alpha\sum_nA_{22}(n)e^{-i\epsilon_n\tau}\approx\frac{\sqrt{k}}{304}&\left\{\cos (J\tau) \left[16 J\tau(3 k+10)+\frac{304}{\sqrt{k}}-3 \frac{J^{\prime 2}\tau}{J} k^2 \right]\right.\nonumber\\
  +&\left.3 \sin (J\tau) \left[48-16k+\frac{J^{\prime 2}\tau}{J} k^{3/2} +\frac{J^{\prime 2}}{J^2} k^2\right]\right\}.
  \label{eq:a22sum_simplified}
\end{align}
The comparison between Eq.~\eqref{eq:a22sum_simplified} and WFMC results are shown in Fig.~\ref{fig:g2tau_WCPB_modified}.

In the short timescales $J\tau\ll 1$,
\begin{align}
  2\alpha\sum_nA_{22}(n)e^{-i\epsilon_n\tau}\approx 1+\sqrt{k}J\tau+\frac{-152+3J^{\prime 2}k^2/J^2}{304}J^2\tau^2
\end{align}
The correlation function $g^{(2)}_{22}(\tau)$ is then given by Eq.~\eqref{eq:g2ijtau_rabi1} with $z$ given by $z=z_0+\delta z$. In the short timescales $J\tau\ll 1$, $g^{(2)}_{22}(\tau)$ behaves as
\begin{align}
  g^{(2)}_{22}(\tau)&=\left|1-2\alpha e^{-iz\tau} \sum_n A_{22}(n)e^{-i\epsilon_n\tau}\right|^2\nonumber\\
 &\approx\left|\frac{(1+i)\sqrt{19}}{8\sqrt{2}}\sqrt{\frac{\alpha}{J}}k^{1/4}J\tau+\frac{1}{128}\left(64k-19i\frac{\alpha}{J}  \sqrt{k}\right)J^2\tau^2 \right|^2.
\end{align}
If we assume negligible nonlinearity $\alpha/J\ll1$ and thus $\gamma=-2\Im{z}\approx-2\Im{z_0}=2\sqrt{k}J$, the result can be further approximated as
\begin{equation}
  g^{(2)}_{22}(\tau)\approx \left|\frac{1}{2}k J^2 \tau^2\right|^2\approx (\gamma \tau)^4/64.
\end{equation}

\begin{figure}
  \centering
  \includegraphics[width=0.4
  \textwidth]{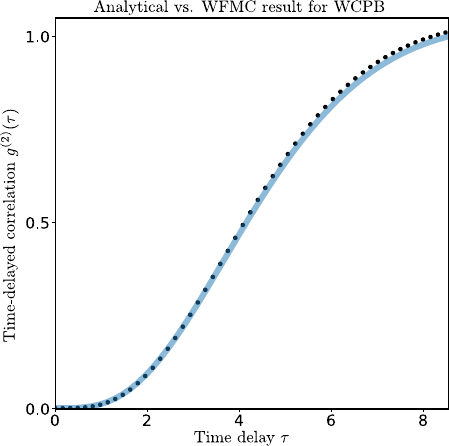}
  \caption{Time-delayed correlation function $g_{22}^{(2)}(\tau)$ for weak-coupling photon blockade in a four-cavity system, using the full analytical formula \eqref{eq:a22sum_simplified} (solid blue curve), and WFMC simulations (black dash).  The system parameters are the same as in Fig.~2 of the main text.} 
  \label{fig:g2tau_WCPB_modified}
\end{figure}

\section{Equal-time second order correlation vs.~photon occupation number}
\label{sec:occupation}

This section looks into the dependence of the equal-time second order correlation on photon occupation in the signal cavity. Due to the weak coupling $J'/k$ between the driven cavity and signal cavity, the photon number in the driven cavity is several orders of magnitude larger than in the signal cavity, making the WFMC simulation with high signal cavity photon occupation difficult.

To overcome this problem and demonstrate the $g^{(2)}(\tau)$ dependence on photon occupation $n$, we use another set of parameters with $k=4$ instead of $k=16$, and $J'=J=0.5386$ instead of $J'=0.2J$.  The other parameters are $\gamma=1$, $\alpha=0.0194$ and $\Delta=0.0652$.  Fig.~\ref{fig:occupation}(a) shows the resulting comparison between the LLPB system, the two-cavity UPB model (with the same $\gamma$ and $\alpha$), and a conventional photon blockade system (with the same $\gamma$ and large $\alpha=10$).  The enlarged antibunching time window and the flat bottom near $\tau=0$ can still be observed. Fig.~\ref{fig:occupation}(b) then shows the equal-time second order correlation versus the photon occupation number in the signal cavity.

\begin{figure}
  \centering
  \includegraphics[width=0.8
  \textwidth]{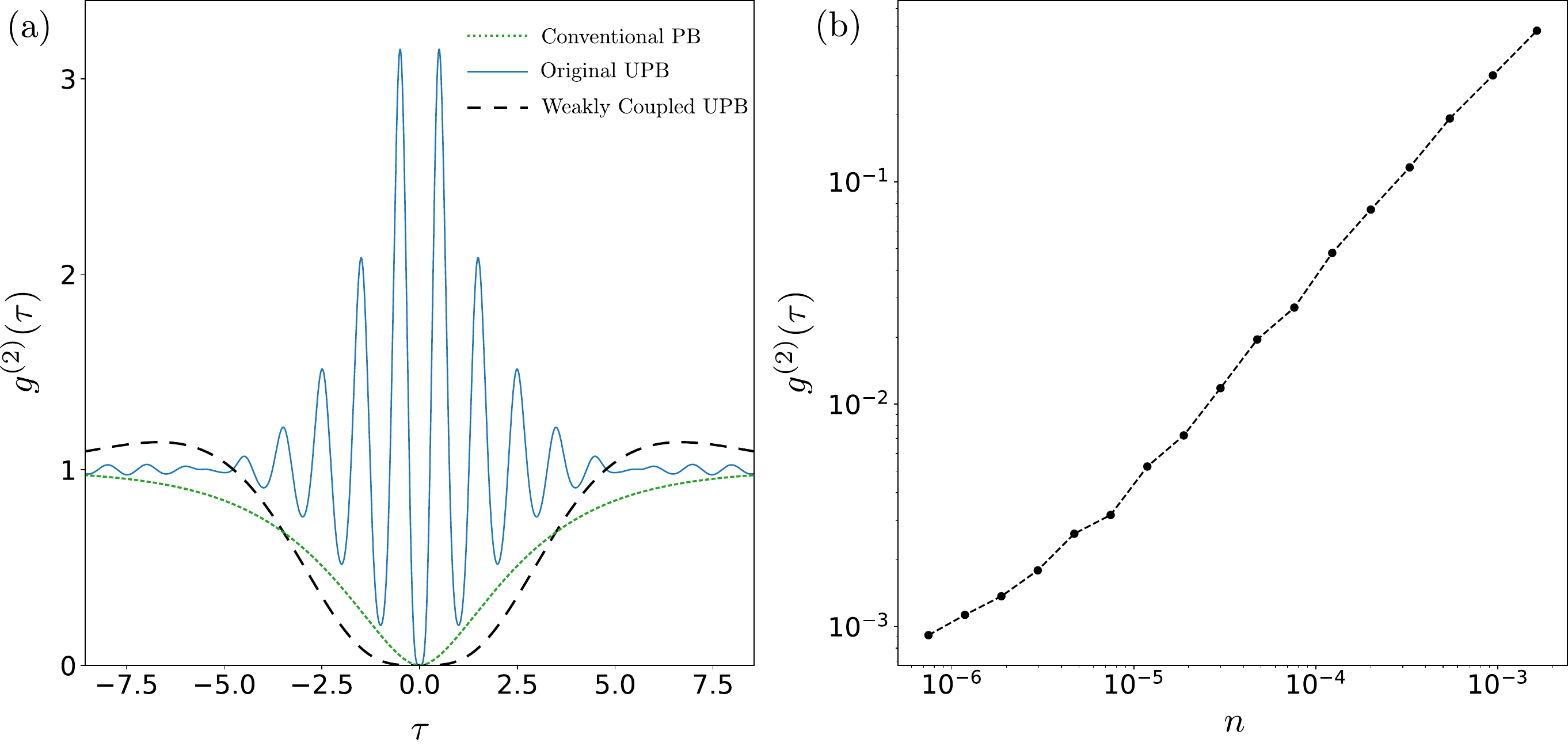}
  \caption{WFMC simulations of the weakly coupled UPB model for another parameter setup with $k=4$, $\gamma=1$, $J'=J=0.5386$, $\alpha=0.0194$ and $\Delta=0.0652$. (a) Comparison with conventional photon blockade and original UPB models. (b) Signal cavity photon occupation number versus equal-time second order correlation.} 
  \label{fig:occupation}
\end{figure}

\section{WFMC simulation details}
Wave function Monte Carlo (WFMC) simulations \cite{Dum92,Molmer93,carmichael_1993,Barchielli_1991} were performed using the \texttt{QuantumOptics.jl} toolbox \cite{kramer2018quantumoptics}, operating in the tensor product of truncated Fock spaces across lattice sites. Due to weak driving and uneven photon occupations, we adopted the following strategies to ensure accuracy and efficiency:

\begin{itemize}
\item To avoid undersampling rare but potentially important jump events, we modified the standard master equation. For jump operators $L_j=\sqrt{\gamma}a_j$, we applied the transformation:
\begin{align}
\Hami &\rightarrow \Hami^{\prime} = \Hami + \frac{\beta}{2i} \sum_j (L_j - L_j^{\dagger}) \\
L_j &\rightarrow L^{\prime}_j = L_j + \beta \mathcal{I}
\end{align}
with $\beta=0.1$. This increases jump rates while preserving the correct ensemble evolution and is known as choosing a different unraveling \cite{BreuerBookOpen}.

\item Simulations were performed with tolerances $\texttt{RelTol} = \texttt{AbsTol} = 10^{-21}$.

\item To capture steady-state behavior, each trajectory was evolved for a relaxation time $T_\text{relax}$, followed by a recording period $T_\text{record}$. We used $T_\text{relax} = 100$, $T_\text{record} = 1000$, and the number of trajectories $N_\text{traj} = 10$.

\item Fock cutoffs $N_\text{Fock}^{(j)}$ were chosen per site to balance accuracy and efficiency. For the driven cavity, we used $N_\text{Fock}^{(1)} = 16$, while for the other cavities, we used $N_\text{Fock}^{(j\neq 1)} = 8$.
\end{itemize}

Convergence was verified in all cases.

\section{Photonics Realization}

In this section, we explore how the quantum optical model could be implemented in practice.  To achieve the four coupled optical modes depicted in Fig.~1(b) of the main text, we can employ either four coupled microring resonators, as shown in Fig.~\ref{fig:rings_gap}(a), or two coupled microring resonators with clockwise (CW) and counter-clockwise (CCW) modes, as shown in Fig.~\ref{fig:rings_gap}(b) and Fig.~1(c) of the main text.  We will focus on the latter case.

\begin{figure}[htbp]
    \centering
    \includegraphics[width=\textwidth]{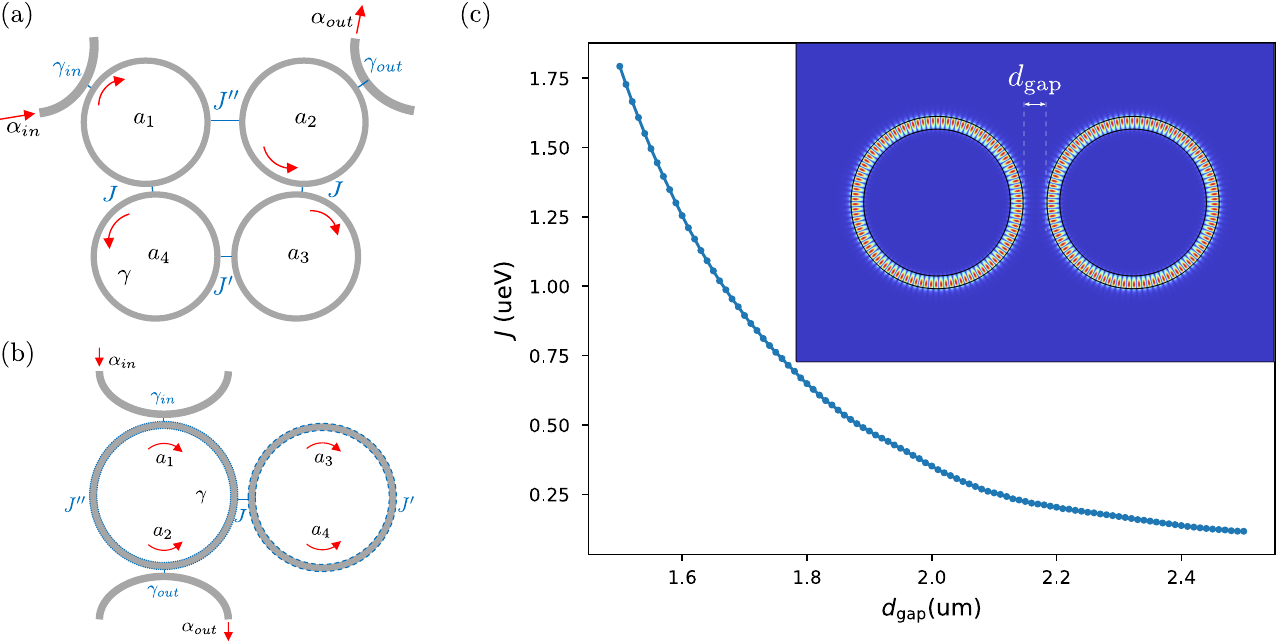}
    \caption{Photonic implementations of the model. (a) Configuration using four coupled microring resonators. (b) Configuration using two coupled microring resonators where the clockwise (CW) and counter-clockwise (CCW) modes within each ring are coupled. (c) Inter-ring coupling strength $J$ as a function of the gap size $d_{gap}$ between the two rings, calculated using Finite Element Method (FEM) simulations.}
    \label{fig:rings_gap}
\end{figure}

In the two-ring setup, the CW and CCW modes have the same natural frequencies.  They are coupled to each other, with coupling rates $J'$ and $J''$, which can be achieved by deliberately-introduced subwavelength defects or surface roughness, as shown in previous studies~\cite{Little97,Morichetti10,Li12,Li20}.  Assuming the two rings have the same shape and size, the Langevin equation for the system is given by
\begin{equation}
i\frac{d}{dt}\begin{pmatrix}
a_1 \\
a_2 \\
a_3 \\
a_4
\end{pmatrix} = 
\begin{pmatrix}
\Delta-\frac{i}{2}(\gamma+\gamma_{in}+\gamma_{out}) & J'' & 0 & J \\
J'' & \Delta-\frac{i}{2}(\gamma+\gamma_{in}+\gamma_{out}) & J & 0 \\
0 & J & \Delta-\frac{i}{2}\gamma & J' \\
J & 0 & J' & \Delta-\frac{i}{2}\gamma
\end{pmatrix}
\begin{pmatrix}
a_1 \\
a_2 \\
a_3 \\
a_4
\end{pmatrix} + 
\begin{pmatrix}
i\sqrt{\gamma_{\textrm{in}}}\alpha_{\textrm{in}} \\
0 \\
0 \\
0
\end{pmatrix},
\end{equation}
where $\alpha_{\textrm{in}}$ is the input field amplitude, $\gamma$ is the intrinsic decay rate of the rings, $\gamma_{\textrm{in}}$ is the input coupling rate, and $\gamma_{\textrm{out}}$ is the output coupling rate. The output field is given by $\alpha_{\textrm{out}}=\sqrt{\gamma_{\textrm{out}}}a_2$.
The parameters $J$, $J'$, and $J''$ represent the coupling strength between the two rings, the coupling strength between the CW and CCW modes within the second ring, and the coupling strength between the CW and CCW modes within the first ring, respectively.
The detuning $\Delta = \omega_c - \omega_L$ is the difference between the cavity resonance frequency $\omega_c$ and the laser frequency $\omega_L$. (We set $\hbar=1$.)

The Hamiltonian terms describing the self-Kerr and cross-Kerr nonlinearities (between CW and CCW modes) are
\begin{equation}
\mathcal{H}_{nl} = \sum_i \alpha a_i^{\dagger}a_i^{\dagger}a_i a_i + 2\alpha a_1^{\dagger}a_1 a_2^{\dagger}a_2 + 2\alpha a_3^{\dagger}a_3 a_4^{\dagger}a_4,
\end{equation}
where $\alpha$ is the Kerr coefficient of each ring resonator.

We perform Finite Element Method (FEM) simulations of this setup, using exemplary parameters drawn from the experimental literature.  We assume each ring is composed of silicon carbide (SiC), with radius $R=3\,\mu$m, waveguide thickness $t=0.35$ $\mu$m, and waveguide width $w=0.8\,\mu$m, operating at a wavelength near $\lambda=1550$ nm. The refractive indices are $n_{\textrm{SiC}} =2.45$ for the waveguides and $n(\text{SiO}_2)=1.44$ for the cladding.  The nonlinear refractive index for SiC is $n_2 = 4.8 \times 10^{-6}$ $\mu$m$^2$/W. The nonlinearity parameter $\alpha$ is estimated by~\cite{Ferretti12,Flayac15}
\begin{equation}
\alpha=\frac{c\hbar^2\omega_c^2}{n_{\textrm{SiC}}^2V}n_2,
\end{equation}
where $V$ is the mode volume, approximated here by the volume of the microring resonator $V=2\pi Rwt$. This yields $\alpha \approx 4.7 \times 10^{-6}$ $\mu$eV.

Figure~\ref{fig:rings_gap}(c) shows the coupling strength $J$ as a function of the gap size $d_{\textrm{gap}}$ between the two rings, derived from FEM simulations. In these simulations, the eigenenergy splitting between the symmetric and antisymmetric supermodes of the coupled system corresponds to $2J$, which is the eigenvalue difference of the matrix
\begin{equation*}
  \begin{pmatrix} \Delta&J\\J&\Delta\end{pmatrix}.
\end{equation*}
The FEM simulation identifies an eigenmode at $\lambda=1553$ nm with a simulated intrinsic quality factor $Q_{\textrm{int}} \approx 1.5\times10^6$. Accounting for other potential loss sources, we estimate the total intrinsic loss rate of the cavity as $\gamma=5$ $\mu$eV, corresponding to an unloaded quality factor $Q_{\textrm{cav}} = \hbar\omega_c / \gamma \approx 1.6\times10^5$. The input and output coupling rates are chosen to be $\gamma_{\textrm{in}}=\gamma_{\textrm{out}}=2.5$ $\mu$eV. The couplings strengths $J'$ and $J''$, can be controlled either by adjusting the gap size $d_{\textrm{gap}}$ in the four-cavity configuration [Fig.~\ref{fig:rings_gap}(a)] or by introducing a subwavelegnth defect or slightly increase surface roughness~\cite{Little97, Li12, Li20, Morichetti10}, that weakly couple the CW and CCW modes in each ring for the two-cavity setup (Fig.~\ref{fig:rings_gap}(b)).  To achieve LLPB, we require $J''\ll J'$, but their actual values do not need to be precisely controlled; thus, we take $J'=0.2$ $\mu$eV and $J''=0.02$ $\mu$eV.

Since experimental parameters may deviate from these nominal values, the detuning $\Delta = \hbar(\omega_c - \omega_L)$ and the inter-ring coupling $J$ are treated as tunable parameters. The detuning $\Delta$ can be adjusted by tuning the laser frequency $\omega_L$. Fine-tuning the coupling $J$ can be achieved by first selecting a gap size $d_{\textrm{gap}}$ corresponding to the desired approximate value of $J$ (from Fig.~\ref{fig:rings_gap}(c)) and then continuously adjusting it. This fine-tuning can be accomplished by illuminating the gap region with a control laser source. Modulating the intensity of this control source allows for precise regional adjustments of the refractive index. Figure~\ref{fig:sim}(b) illustrates the dependence of $J$ on the regional refractive index change $\Delta n$ within a spot of radius $r=0.5$ $\mu$m located in the gap (schematically shown in Fig.~\ref{fig:sim}(a)), assuming an initial gap size of $d_{\textrm{gap}}=1.74$ $\mu$m.

Figure~\ref{fig:sim}(c) displays a colormap of the zero-delay second-order correlation function $g^{(2)}(0)$ as a function of the detuning $\Delta$ and coupling strength $J$. The black contour line indicates the boundary of the antibunching region, defined by $g^{(2)}(0) < 0.5$. The inset shows the time-delayed second-order correlation function $g^{(2)}(\tau)$ for the output field, calculated at the parameter values marked by the purple cross in the main panel. It shows a over $1$ ns time window for $g^{(2)}(\tau)<0.5$.

\begin{figure}[htbp]
    \centering
    \includegraphics[width=\textwidth]{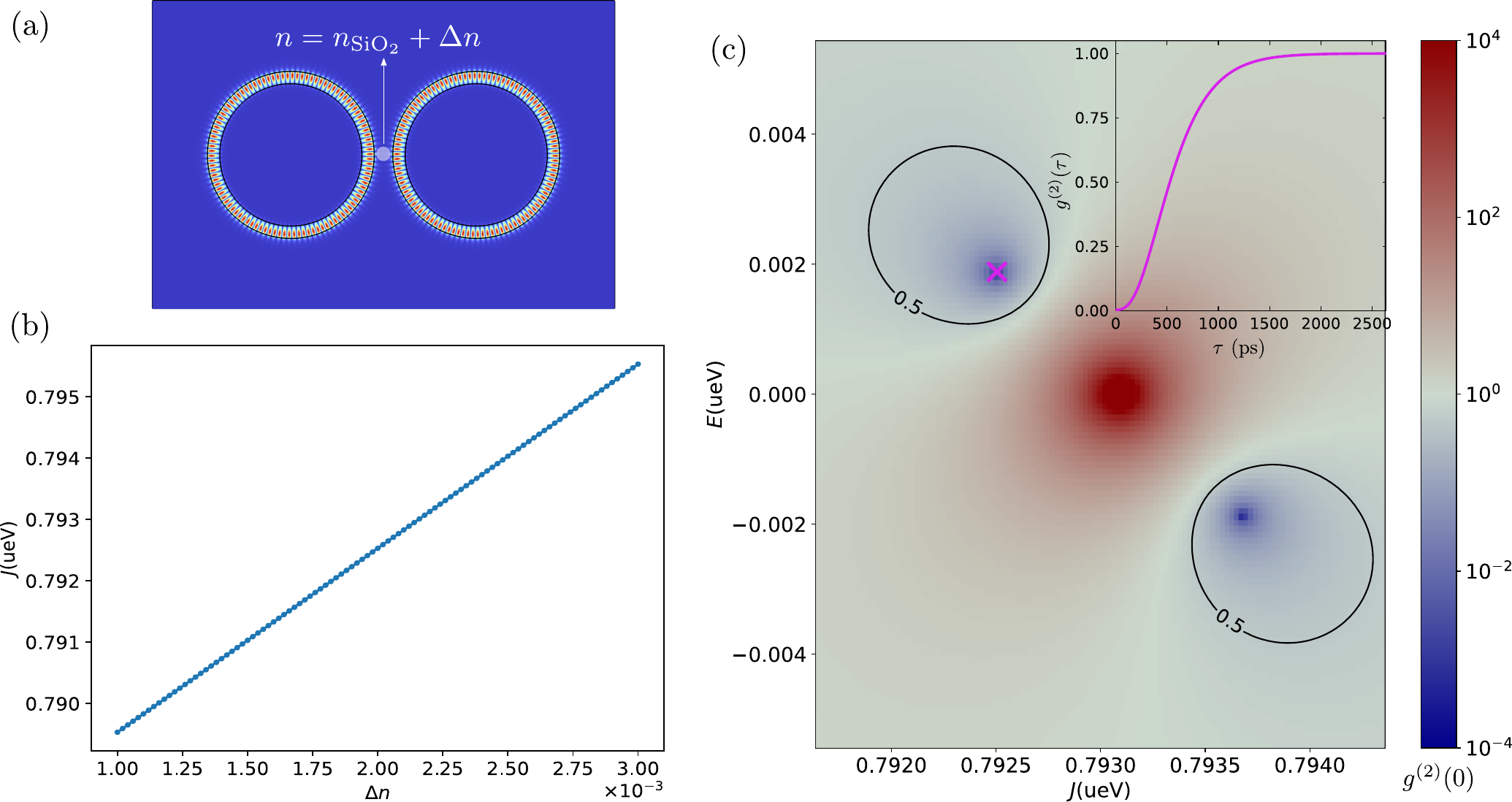}
    \caption{Fine-tuning and simulation results. (a) Schematic illustrating the $r=0.5$ $\mu$m spot within the inter-ring gap used for refractive index-based fine-tuning of the coupling $J$. (b) Dependence of the coupling strength $J$ on the regional refractive index change $\Delta n$ applied to the spot shown in (a), assuming a fixed gap size $d_{\textrm{gap}}=1.74$ $\mu$m. (c) Colormap of the zero-delay second-order correlation function $g^{(2)}(0)$ for the output field as a function of detuning $\Delta$ and coupling strength $J$. The black contour line ($g^{(2)}(0) = 0.5$) delineates the antibunching region. The inset displays the time-delayed second-order correlation $g^{(2)}(\tau)$ calculated at the parameter point indicated by the purple cross.}
    \label{fig:sim}
\end{figure}

\clearpage
\end{widetext}

\end{document}